\documentclass[10pt,a4paper]{article}
\usepackage[ascii]{inputenc}
\usepackage{amsmath,amssymb,amsfonts,amsthm}
\usepackage[a4paper,lmargin=3.9cm,rmargin=3.9cm]{geometry}
\usepackage{cite}
\usepackage{graphicx}
\usepackage{upquote}
\usepackage[caption=false]{subfig}
\usepackage[pdftex,bookmarks=false,colorlinks=true,linkcolor=black,
citecolor=black,filecolor=black,urlcolor=blue]{hyperref}

\graphicspath{{figures/}}

\begin{document}

\title{GuiTeNet: A graphical user interface for \\ tensor networks}

\author{Lisa Sahlmann and Christian B. Mendl\footnote{Technische Universit\"at Dresden, Institute of Scientific Computing, Zellescher Weg 12-14, 01069 Dresden, Germany; \href{mailto:lisa.sahlmann@tu-dresden.de}{lisa.sahlmann@tu-dresden.de}, \href{mailto:christian.mendl@tu-dresden.de}{christian.mendl@tu-dresden.de}}}

\date{July 30, 2018}

\maketitle

\begin{abstract}
We introduce a graphical user interface for constructing arbitrary tensor networks and specifying common operations like contractions or splitting, denoted GuiTeNet. Tensors are represented as nodes with attached legs, corresponding to the ordered dimensions of the tensor. GuiTeNet visualizes the current network, and instantly generates Python/NumPy source code for the hitherto sequence of user actions. Support for additional programming languages is planned for the future. We discuss the elementary operations on tensor networks used by GuiTeNet, together with high-level optimization strategies. The software runs directly in web browsers and is available online at \href{http://guitenet.org}{guitenet.org}.
\end{abstract}

\section{Introduction}

Tensor networks have found a wide range of applications within mathematics \cite{HackbuschKuehn2009, Hackbusch2014}, physics and chemistry, in particular as matrix product states (MPS), projected entangled pair states (PEPS) or the multiscale entanglement renormalization ansatz (MERA) for strongly correlated quantum systems \cite{Schollwock2011, VerstraeteMurgCirac2008, Vidal2008}. While tensor networks and associated operations are conveniently represented as graphical diagrams, a subsequent implementation of these operations is often tedious, especially if one has to keep track of arrangements of many indices. On the other hand, fundamental operations like finding the (quasi-)optimal contraction order and performing the partial contraction of a given tensor network are available as software packages \cite{NCON2014, PfeiferHaegemanVerstraetePRE2014, EvenblyPfeiferPRB2014} or via NumPy's \verb|einsum| command \cite{WaltColbertVaroquaux2011, NumpyEinsum}. To bridge the gap between graphical representation and implementation, we introduce a graphical user interface (GUI) for constructing arbitrary tensor networks and specifying common operations on them, like contractions or splitting via QR- or SVD-decompositions. Our software framework then instantly generates source code for these operations; currently Python/NumPy is supported, with additional programming languages planned for the future. We use JavaScript and the D3.js library to make the GUI conveniently available via web browsers.

\section{Description of the GUI}
\label{sec:description_interface}

\begin{figure}[!ht]
\centering
\includegraphics[width=0.225\textwidth]{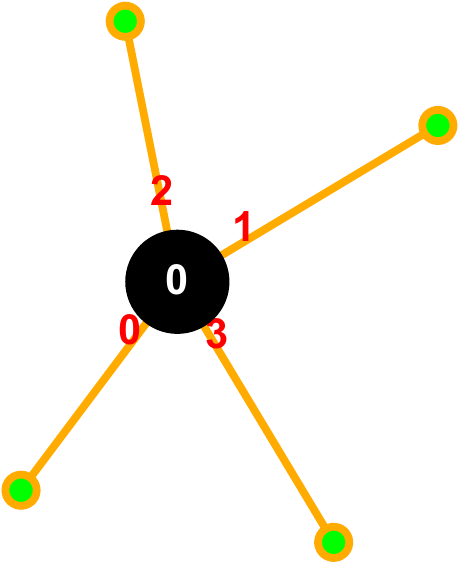}
\caption{A single tensor with 4 legs (dimensions). The ordering of dimensions is indicated by the red labels.}
\label{fig:tensor}
\end{figure}

The GUI represents each tensor as a node with an arbitrary number of legs, corresponding to the number of dimensions (rank) of the tensor. The ordering of the dimensions is indicated by labels. Fig.~\ref{fig:tensor} visualizes a single tensor as it appears in the GUI. Note that this abstract representation leaves the actual dimensions open, i.e., it does not differentiate between, say, a $2 \times 3 \times 4$ tensor and a $8 \times 7 \times 6$ tensor, since both have rank $3$.

The user interacts with the GUI mainly via drag-and-drop gestures, to add tensors to the network or attach legs to a tensor, and to specify operations like contractions and tensor splitting; see below for more details. The GuiTeNet framework visualizes the current tensor network, and simultaneously generates source code which implements the hitherto sequence of user actions. For example, the generated Python code for a contraction of three tensors followed by QR splitting reads:
{\small
\begin{verbatim}
import numpy as np

def f(T0, T1, T2):
    T3 = np.einsum(T0, (0, 1, 2), T1, (3, 2), T2, (0, 4, 5), (1, 3, 4, 5))
    T4 = np.transpose(T3, (3, 0, 2, 1))
    T5, T6 = np.linalg.qr(T4.reshape((np.prod(T4.shape[:2]),
                  np.prod(T4.shape[2:]))), mode='reduced')
    T5 = T5.reshape(T4.shape[:2] + (T5.shape[1],))
    T6 = T6.reshape((T6.shape[0],) + T4.shape[2:])
    return (T5, T6)
\end{verbatim}
}
\noindent Details of the code generation are provided in section~\ref{sec:elementary_operations}.

\paragraph{Creating tensors}

\begin{figure}[!ht]
\centering
\subfloat[select]{\includegraphics[width=0.225\textwidth]{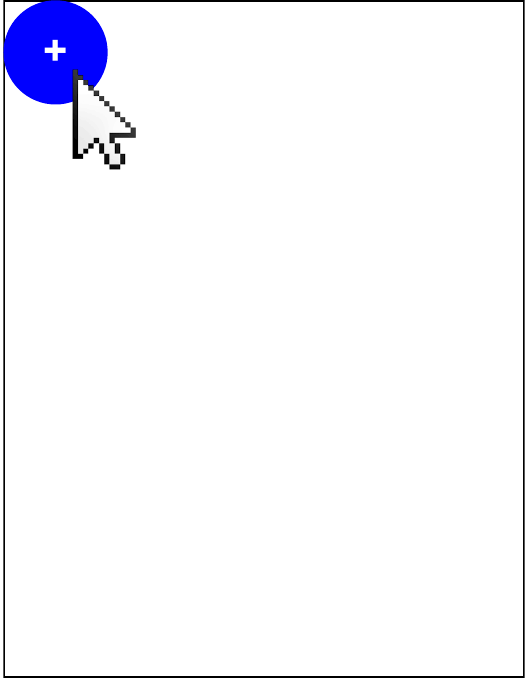}}%
\hspace{0.02\textwidth}%
\raisebox{1.6cm}{$\to$}%
\hspace{0.02\textwidth}%
\subfloat[drag]{\includegraphics[width=0.225\textwidth]{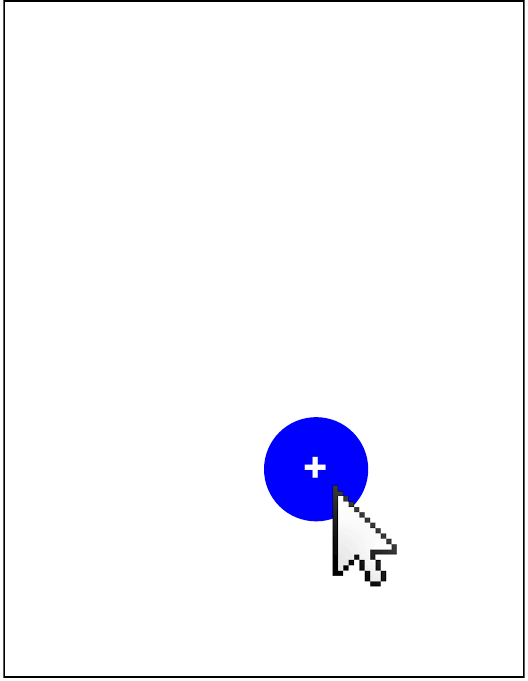}}%
\hspace{0.02\textwidth}%
\raisebox{1.6cm}{$\to$}%
\hspace{0.02\textwidth}%
\subfloat[drop]{\includegraphics[width=0.225\textwidth]{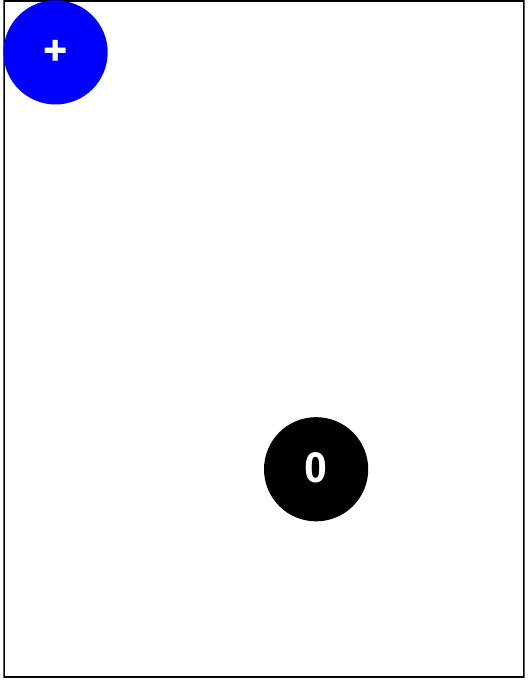}}%
\caption{Creating a new tensor by a drag-and-drop gesture. The mouse pointer is enlarged for visual clarity.}
\label{fig:create_tensor}
\end{figure}

A new tensor is added to the network by a drag-and-drop gesture. The user drags a special ``create tensor'' symbol (blue circle in Fig.~\ref{fig:create_tensor}) to the desired location. When ``dropping'' the symbol, a new tensor (black circle) appears there. Initially it has zero legs. The tensors are automatically labeled $0, 1, 2, \dots$ to provide a unique identifier. The ``create tensor'' symbol reappears at its default location after this operation, and can then be used to add another tensor to the network.

\paragraph{Attaching tensor legs}

Each leg represents one dimension of the tensor. The user creates a new leg by ``pulling'' it out of the tensor (i.e., drag-and-drop on the tensor), when simultaneously holding the Control key. Each tensor and its legs can still be freely moved around within the GUI window.

\paragraph{Tensor contractions}

\begin{figure}[!ht]
\centering
\subfloat[input network]{\includegraphics[width=0.45\textwidth]{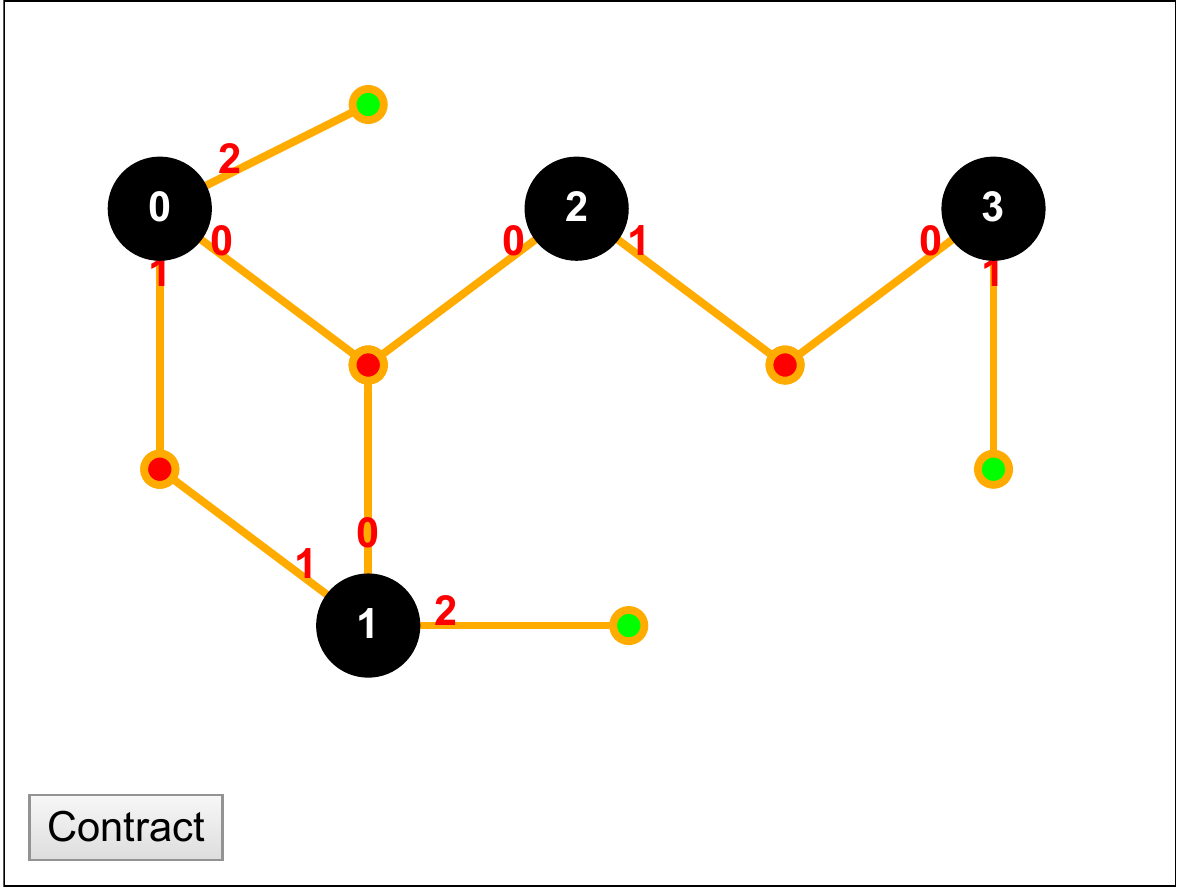}}%
\hspace{0.02\textwidth}%
\raisebox{1.9cm}{$\to$}%
\hspace{0.02\textwidth}%
\subfloat[output tensor]{\includegraphics[width=0.45\textwidth]{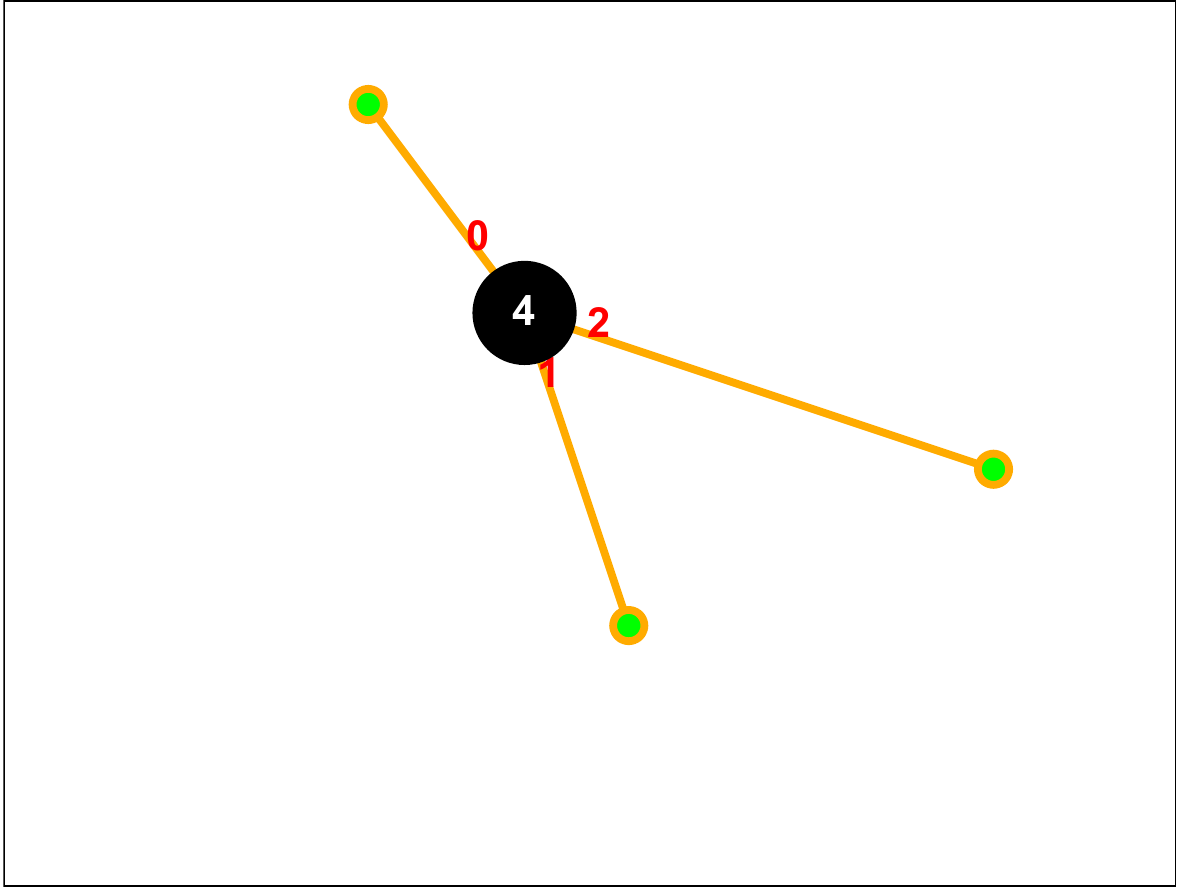}}%
\caption{Illustration of an elementary contraction operation.}
\label{fig:contraction}
\end{figure}

Tensor contractions are specified by connecting the tips of tensor legs. The tips snap to each other when brought into close contact. The actual contraction (possibly of several tensors) is executed when pressing the ``Contract'' button of the GUI, see Fig.~\ref{fig:contraction} for an example.

\paragraph{Splitting a tensor}

\begin{figure}[!ht]
\centering
\subfloat[input tensor]{\includegraphics[width=0.3\textwidth]{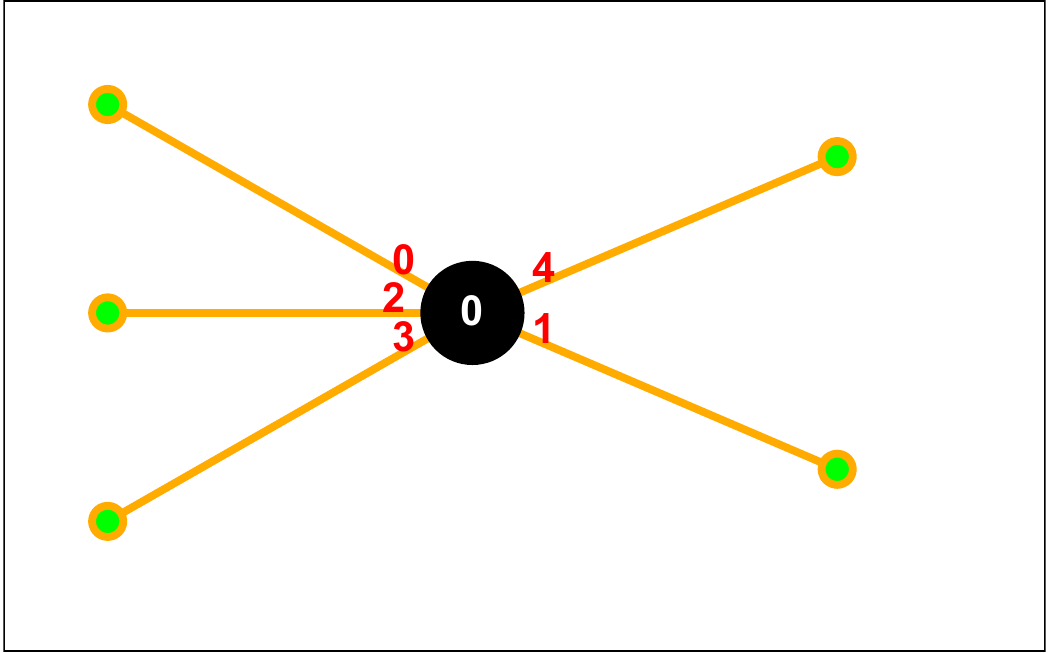}}%
\hspace{0.01\textwidth}%
\raisebox{1.1cm}{$\to$}%
\hspace{0.01\textwidth}%
\subfloat[splitting parameters]{\includegraphics[width=0.3\textwidth]{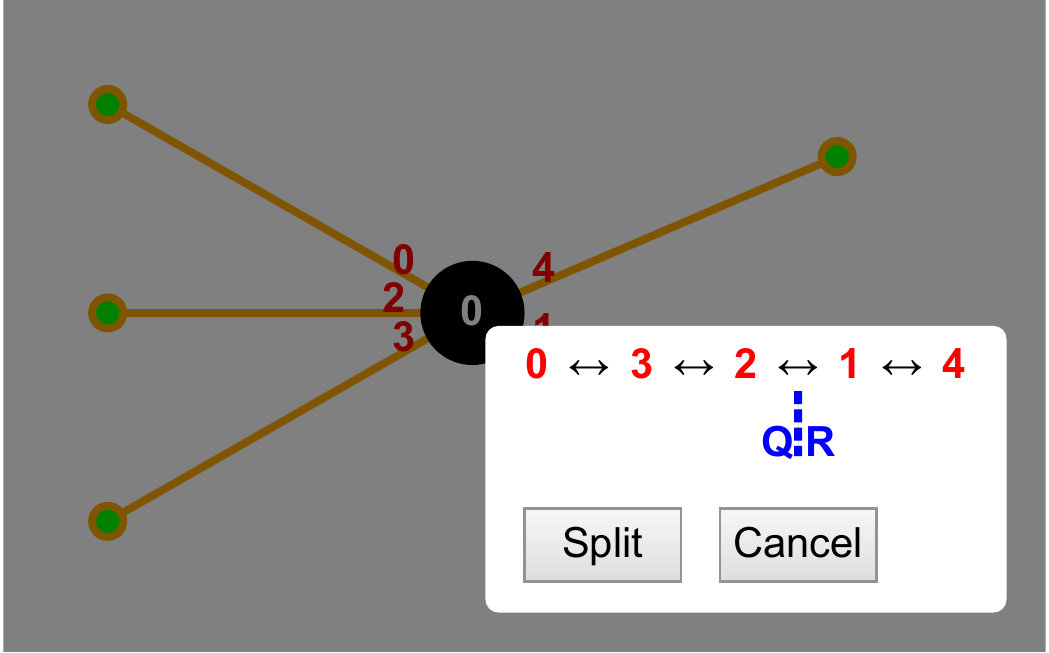}}%
\hspace{0.01\textwidth}%
\raisebox{1.1cm}{$\to$}%
\hspace{0.01\textwidth}%
\subfloat[output $Q$ and $R$ tensors]{\includegraphics[width=0.3\textwidth]{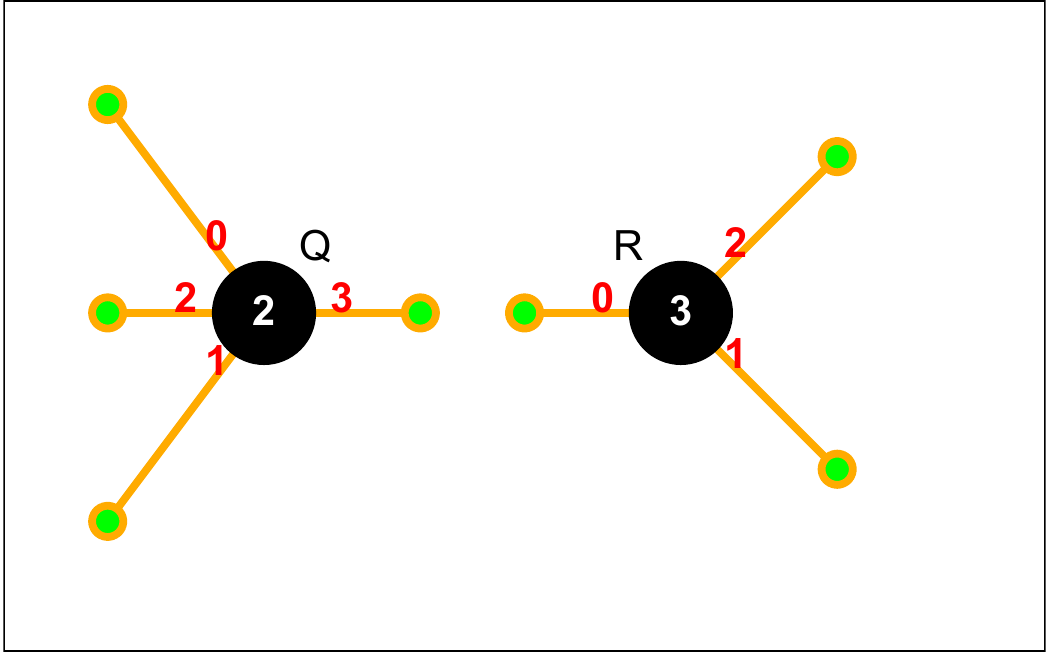}}%
\caption{QR splitting of a tensor. In (b), the user specifies the ordered dimensions attributed to the $Q$ and $R$ tensors, respectively.}
\label{fig:qr_splitting}
\end{figure}

The splitting of a tensor by QR or singular value decomposition (SVD) is a ubiquitous operation in tensor network algorithms, in particular for reducing ``bond dimensions'' by devising a singular value cut-off tolerance, and a prerequisite for working with left- and right-orthogonal tensors in the MPS framework \cite{Schollwock2011}. The first step for decomposing a tensor $A$ is its ``matricization'': a subset of legs is grouped together into one ``fat'' leg and the remaining (complementary) legs into a second ``fat'' leg. The two fat legs are interpreted as the rows and columns of a matrix, which is then decomposed. Fig.~\ref{fig:qr_splitting} illustrates this process (as it appears in the GUI) for the QR decomposition of a tensor with initially 5 legs. (An analogous SVD decomposition is currently still under development.) The user first right-clicks on a tensor to initiate the splitting operation. An overlay window then asks for the ordering and partitioning of dimensions attributed to the rows and columns in the matricization process. In the example, the ``row'' consists of dimensions $0$, $3$, $2$ (in this order) and the ``column'' of dimensions $1$, $4$ (in this order). After the decomposition, the resulting $Q$ and $R$ matrices are finally reshaped to restore the original dimensions, with an additional dimension for the shared bond (last dimension of $Q$, first dimension of $R$). Thus the dimensions $0$, $1$, $2$ of $Q$ match the original dimensions $0$, $3$, $2$, and dimensions $1$, $2$ of $R$ the original dimensions $1$, $4$.

The initial reordering of dimensions becomes a separate ``elementary transposition operation'', as described in section~\ref{sec:elementary_operations} below. The generated code uses a temporary tensor for this purpose. In Fig.~\ref{fig:qr_splitting}, this temporary tensor has index $1$, and hence the $Q$ and $R$ tensors are consecutively labeled $2$ and $3$.

After this reordering, the partitioning is simply a reinterpretation of the data stored in the tensor, since the ``row'' group now consists of the first $\ell$ leading dimensions, and the ``column'' group of the remaining $r - \ell$ trailing dimensions, where the rank $r$ is the total number of dimensions.

\section{Elementary tensor network operations}
\label{sec:elementary_operations}

Somewhat analogous to an intermediate representation in source code compilation, we decompose the actions supported by the GUI into the following elementary operations on tensor networks:

\paragraph{(i) Elementary contraction of tensors} The GuiTeNet framework supports general contraction operations on a tensor network. An \emph{elementary} contraction acts on a subset of tensors such that these tensors are joined (directly or indirectly) by shared legs, yielding a single tensor after the contraction. Note that ``multi-bond'' contractions, i.e., the simultaneous contraction of multiple legs as in Fig.~\ref{fig:contraction}, is explicitly allowed. In principle, a contraction of several tensors could be decomposed into a sequence of pairwise contractions, e.g., computing the matrix product $A B C$ by first ``contracting'' $A$ with $B$ to obtain $T = A B$ and then multiplying $T$ with $C$. However, in general the optimal order of these pairwise contractions poses a delicate optimization problem \cite{PfeiferHaegemanVerstraetePRE2014} and is not straightforwardly applicable to multi-bond contractions. Hence we regard the contraction of (possibly more than two) tensors as elementary operation, and leave the optimized implementation to backend software packages.

On the other hand, a sequence of tensor network operations can be optimized by merging subsequent elementary contractions into a single elementary contraction. As simple (toy model) illustration why this might be useful, consider the contraction $C = A B$ (matrix-matrix multiplication) followed by the contraction $y = C x$ (matrix-vector product). Merging these two contractions leads to $y = A B x$, for which a backend algorithm would naturally choose the order $y = A (B x)$.

To uniquely specify a contraction operation, we follow NumPy's \verb|einsum| command convention in the form \verb|einsum|$(T_0, s_0, T_1, s_1, \dots, s_{\text{out}})$. Here the $T_i$ refer to tensors, and $s_i$ are lists of integer labels for the corresponding dimensions, with multiply occurring labels to be summed over. The last argument $s_{\text{out}}$ determines the ordering of dimensions in the output tensor after the contraction. For the example in Fig.~\ref{fig:contraction} with $4$ tensors,
\begin{equation*}
s_0 = (0, 1, 2), \quad s_1 = (0, 1, 3), \quad s_2 = (0, 4), \quad s_3 = (4, 5) \quad \text{and} \quad s_{\text{out}} = (2, 3, 5).
\end{equation*}
Thus, the three dimensions of tensor $T_0$ are labeled $0$, $1$, $2$, the three dimensions of tensor $T_1$ are labeled $0$, $1$, $3$ etc. The dimensions labeled $0$, $1$ and $4$ will be contracted since they appear multiple times, and the remaining dimensions are ordered as $(2, 3, 5)$ in the output tensor. The generated Python source code follows exactly this scheme and reads explicitly
{\small
\begin{verbatim}
T4 = np.einsum(T0, (0, 1, 2), T1, (0, 1, 3),
               T2, (0, 4), T3, (4, 5), (2, 3, 5))
\end{verbatim}
}

\paragraph{(ii) General transposition of a tensor} Formally, a tensor transposition is a permutation of dimensions, generalizing the usual transposition of matrices. For example, applying the permutation $(1, 2, 0)$ to a $10 \times 11 \times 12$ tensor $T$ yields a $11 \times 12 \times 10$ tensor, such that the $(i,j,k)$-th entry of $T$ is the $(j,k,i)$-th entry of the transposed tensor. Since the tensor elements are typically stored as a contiguous array in memory, a transposition implies a reshuffling of the array elements. Thus, while a transposition does not involve arithmetic calculations besides computing memory addresses, its cost can still be significant, in particular due to the inherent ``cache-unfriendliness''.

Specifying a transposition only requires designating the permutation of dimensions. We follow the convention of NumPy's \verb|transpose| function.

Regarding transpositions as separate elementary operations --- instead of first step for splitting a tensor for example --- facilitates additional optimizations. A plausible scenario is integrating the transposition into a preceding contraction operation \cite{Matthews2018}, generalizing $(A B^T)^T = B A^T$ for matrices $A$ and $B$.

\paragraph{(iii) QR decomposition of a tensor} The elementary QR decomposition considered here does not involve any reordering of dimensions. Thus, as described in section~\ref{sec:description_interface}, it is uniquely specified by the number $\ell$ of leading tensor dimensions to be interpreted as ``row'' dimension in the matricization process, and correspondingly the remaining dimensions as ``column'' dimension.

To illustrate, the generated Python code (up to renaming variables) for the elementary QR decomposition of a tensor $T$ and $\ell = 3$ leading dimensions reads as follows:
{\small
\begin{verbatim}
Q, R = np.linalg.qr(T.reshape((np.prod(T.shape[:3]), np.prod(T.shape[3:]))),
                    mode='reduced')
Q = Q.reshape(T.shape[:3] + (Q.shape[1],))
R = R.reshape((R.shape[0],) + T.shape[3:])
\end{verbatim}
}
\noindent The \verb|reshape| functions implement the matricization before and ``de-matricization'' after the actual QR decomposition, \verb|T.shape| stores the tensor dimensions, \verb|np.prod| computes the product of the leading and trailing dimensions, and \verb|np.linalg.qr| implements the conventional QR decomposition of matrices.

\paragraph{(iv) Singular-value decomposition of a tensor} The (de-)matricization process for an elementary singular-value decomposition (SVD) of a tensor is analogous to the elementary QR decomposition. The output now consists of three tensors, corresponding to the $U$, $S$ and $V$ matrices of the matrix-SVD, with the diagonal $S$ matrix storing the singular values. Additional parameters (compared to the QR decomposition) are a cut-off tolerance for the singular values, and optionally the maximally allowed number of singular values (the maximal ``bond'' dimension).

\section{Strategies for optimization}

Based on the elementary tensor network operations, several high-level optimization strategies are conceivable, solely based on the rank of each tensor instead of the actual dimensions.

A natural representation for the sequence of user actions is a directed acyclic graph (DAG), storing an elementary operation or input tensor at each node. Such a representation clarifies dependencies, and allows to determine which operations can be executed in parallel.

A more subtle optimization strategy tailored to tensor networks is the merging of subsequent contractions, i.e., if the tensor resulting from a contraction is immediately used in another contraction. A toy model example (as mentioned in section~\ref{sec:elementary_operations}) consists of merging $C = A B$ followed by $y = C x$ into $y = A B x$, which can then be evaluated in the order $y = A (B x)$. Note that the optimized computational cost for the merged contraction cannot be higher than performing the contractions sequentially (since the latter restricts the allowed contraction order), but actually determining the optimal order for the merged contraction (by a backend software package) is in general more difficult \cite{PfeiferHaegemanVerstraetePRE2014}.

Another optimization strategy is avoiding explicit transpositions (i.e., permutations of tensor dimensions), and aiming for advantageous dimension ordering. As mentioned, the transposition of a tensor resulting from a contraction can be integrated into the contraction (see also \cite{Matthews2018}), generalizing $(A B^T)^T = B A^T$ for matrices. Transpositions may also be pushed through the computational graph; for example, instead of permuting the leading dimensions of the $Q$-tensor resulting from a QR decomposition, one could already permute these dimensions in the input tensor, or vice versa.

\section{Conclusions and outlook}

In its present form, the GuiTeNet software framework is well suited to handle a relatively small number of tensors, but manually constructing a network with hundreds of tensors is cumbersome. Instead, generating code for subroutines or blocks inside loops is a plausible scenario for employing GuiTeNet in larger software projects. As specific example, rather than instantiating all tensors of a matrix product state, the GuiTeNet framework could be used to generate a local contraction operation required during a left-right sweep over the chain.

We also want to point out the pedagogical value which GuiTeNet might offer, including the seamless transition from vectors and matrices to general tensors.

Nevertheless, there are many desirable features left for future work, including code generation for other programming languages and software libraries, or a timeline of previous network states (e.g., before a contraction) with associated \emph{Undo} functionality. Tensors with special properties (like orthogonal tensors resulting from a SVD or QR decomposition) should be marked, e.g., using a different symbol, and ideally such properties should be exploited in the generated code. Furthermore, one could take $U(1)$-symmetries into account by endowing the legs with additive quantum numbers (like particle or spin) and a directional arrow. Conceptually, the sum of quantum numbers flowing into a tensor must be equal to the sum of quantum numbers leaving the tensor, enforcing a block sparsity structure of the tensor. Another worthwhile goal is incorporating more exotic tensor network operations, like ``loop skeletonization'' \cite{Ying2017}.

An interesting open question is how GuiTeNet could inspire or profit from software and hardware architectures tailored to tensor operations, like contractions beyond conventional BLAS routines \cite{Matthews2018} or Google's Tensor Processing Units (TPUs) \cite{GoogleTPU} employed in the TensorFlow machine learning framework.

We encourage active contributions and further development of GuiTeNet, see \href{http://guitenet.org}{guitenet.org} for details.

\paragraph{Acknowledgments} CM likes to thank Lexing Ying for inspiring discussions.

\end{document}